\documentclass[dvipdfmx]{pasj00}
\draft
\Received{2014 March 28}
\Accepted{2014 November 11}
\Published{}
\SetRunningHead{Astronomical Society of Japan}
{Probing Precise Location of Radio Core in TeV Blazar Mrk~501 with VERA at 43~GHz}
\usepackage{times}
\usepackage{natbib}
\usepackage{ulem}
\usepackage{graphicx}
\usepackage{color}
\begin{document}

\title{Probing Precise Location of Radio Core in TeV Blazar Mrk~501\\ with VERA at 43~GHz}
\author{Shoko \textsc{Koyama}\altaffilmark{1,2,3,4},
Motoki \textsc{Kino}\altaffilmark{5,3}, 
Akihiro \textsc{Doi}\altaffilmark{3},
Kotaro \textsc{Niinuma}\altaffilmark{6},
Kazuhiro \textsc{Hada}\altaffilmark{2,7},\\
Hiroshi \textsc{Nagai}\altaffilmark{2},
Mareki \textsc{Honma}\altaffilmark{2},
Kazunori \textsc{Akiyama}\altaffilmark{1,2},
Marcello \textsc{Giroletti}\altaffilmark{7},\\
Gabriele \textsc{Giovannini}\altaffilmark{7,8},
Monica \textsc{Orienti}\altaffilmark{7},
Naoki \textsc{Isobe}\altaffilmark{3},
Jun \textsc{Kataoka}\altaffilmark{9},
David \textsc{Paneque}\altaffilmark{10},
Hideyuki \textsc{Kobayashi}\altaffilmark{2},
and Keiichi \textsc{Asada}\altaffilmark{11}}

\altaffiltext{1}{Department of Astronomy, Graduate School of Science, The University of Tokyo, 7-3-1 Hongo, Bunkyo-ku, Tokyo 113-0033, Japan} 
\altaffiltext{2}{National Astronomical Observatory of Japan, 
2-21-1 Osawa, Mitaka, Tokyo 181-8588, Japan}
\altaffiltext{3}{Institute of Space and Astronautical Science, Japan Aerospace Exploration Agency, 3-1-1 Yoshinodai, Chuou-ku, Sagamihara, Kanagawa 252-5210, Japan}
\altaffiltext{4}{Max-Planck-Institut f\"{u}r Radioastronomie, Auf dem H\"{u}gel 69, D-53121 Bonn, Germany}
\email{skoyama@mpifr-bonn.mpg.de}
\altaffiltext{5}{Korea Astronomy and Space Science Institute, 776 Daedeokdae-ro, Yuseong-gu, Daejeon, 305-348, Republic of Korea}
\altaffiltext{6}{Graduate School of Science and Engineering, Yamaguchi University, Yamaguchi 753-8511, Japan}
\altaffiltext{7}{INAF Istituto di Radioastronomia, via Gobetti 101, 40129 Bologna, Italy}
\altaffiltext{8}{Dipartimento di Astronomia, Universit$\grave{a}$ di Bologna, via Ranzani 1, I-40127 Bologna, Italy}
\altaffiltext{9}{Research Institute for Science and Engineering, Waseda University, 3-4-1, Okubo, Shinjuku, Tokyo 169-8555, Japan}
\altaffiltext{10}{Max-Planck-Institut f\"{u}r Physik, M\"{u}nchen, D-80805, Germany}
\altaffiltext{11}{Institute of Astronomy and Astrophysics, Academia Sinica Institute of Astronomy and Astrophysics, Taipei 10617, Taiwan}


\maketitle

\begin{abstract}
We investigate the position of the radio core in a blazar by multi-epoch
astrometric observations at 43~GHz.
Using the VLBI Exploration of Radio Astrometry (VERA),
we have conducted four adjacent observations 
in February 2011 and another four in October 2011,
and
succeeded in measuring the position of the radio core in the TeV blazar Mrk~501
relative to a distant compact quasar NRAO~512. 
During our observations, we find that
(1) there is no positional change within $\sim0.2~$mas or $\sim2.0~$pc de-projected with $\pm1\sigma$ error
for the weighted-mean phase-referenced positions of Mrk~501 core 
relative to NRAO~512 over four adjacent days, and
(2) there is an indication of position change for 3C~345 core relative to NRAO~512.
By applying our results to the standard internal shock model for blazars,
we constrain the bulk Lorenz factors of the ejecta. 
\end{abstract}

\section{Introduction}
Locating radio-emitting regions in 
relativistic jets in active galactic nuclei (AGNs) 
has been one of the most intriguing issues for 
exploring the ultimate mechanism of jet 
formation \citep[e.g.][]{Marscher:2008,Abdo:2010b}.
One of the leading scenarios 
suggests that radio cores of blazars
correspond to stationary standing shocks located at 
several parsec scales downstream 
from the central black holes,
based on the delay timescales of the radio core 
brightening after the precedent $\gamma$-ray flares
\citep{Marscher:2008,Jorstad:2010,Agudo:2011,Agudo:2011a}.
On the other hand, 
phase-referencing Very Long Baseline Array (VLBA) 
observations of the radio galaxy M87
revealed that 
the radio core at 43~GHz
is located only $\sim 0.01~{\rm pc}$
away from the upstream end of the conical jet
\citep{Hada:2011}.
Nowadays, it is important to explore the origin of
such a discrepancy of radio core locations and their stationarity
between radio galaxies and blazars.

To explore the locations and their stationarity
of radio cores in blazars directly,
multi-epoch Very Long Baseline Interferometry (VLBI) 
astrometric observations are the most powerful and reliable method.
By means of astrometric observations, 
we can measure the positions 
of target sources relative to phase-calibrators close 
to the targets on the sky.  
Over the past few decades, 
astrometric monitoring experiments based 
on the International VLBI Service for Geodesy and Astrometry (IVS) 
also have been performed towards the radio cores of more than a hundred of blazars 
\citep[e.g.,][]{Ma:1998,Fey:2004,Titov:2011}.
However, 
most of these astrometric experiments have been limited 
at 2.3 \& 8~GHz\footnote{see however \cite{Lanyi:2010}}.
Hence, little is known about the relative location of brightness peak and the stationarity
seen at higher frequencies.
%
%

We should carefully select an appropriate 
blazar and an observing frequency.
It is well known that radio cores of blazars usually tend to be optically thick against synchrotron self-absorption (SSA) at lower frequencies \citep[e.g.,][]{Kellermann:1981,Marscher:1985}.
To avoid the SSA effect as much as we can,
we select a TeV blazar,
Mrk~501 ($z = 0.034$). %
Mrk~501 is one of the best blazars for this study
because the radio core spectrum
between 1.4~GHz and 86~GHz measured by VLBI clearly shows 
convex shape peaking at $\sim8~$GHz
\citep[figure~8 in][]{Giroletti:2008}.
Therefore, we can probe the
locations of radio emitting regions
reducing the SSA effect by selecting the observing frequency
higher than the peak frequency. 
We select the observing frequency at 43~GHz because of 
the transparency against SSA.
%
%
The proximity and brightness of Mrk~501 enable us to perform the precise astrometric observations.
Another advantage of this source is that there are two bright phase-calibrators 
within an angular separation of a few degrees, such that
it makes easier to perform phase-referencing observations. 

The plan of this paper is as follows.
In \S2, we describe our astrometric procedure 
for measuring the core position of Mrk~501.
We explain the observation in \S3, and data reduction in \S4. 
%
The results, and the discussions are described in \S5 and \S6.
The conclusion and future prospects are given in \S7.
The error estimation of the phase-referenced positions is written in appendix~\ref{veraerrsec}.
Throughout this paper, we adopt the following 
cosmological parameters;
$H_{0}=71~ {\rm km~s^{-1} Mpc^{-1}}$,
$\Omega_{\rm M}=0.27$, and    
$\Omega_{\rm \Lambda}=0.73$ \citep{Komatsu:2009}, or
$1~{\rm mas}=0.662$~pc for Mrk~501.
%

\section{Astrometric procedure in measuring Mrk~501's core position}~\label{strat}
In this section, we describe our method to
measure the core position of the target source Mrk~501. 
We chose the distant point source NRAO 512
($z=1.66$, $1~{\rm mas}=8.56~{\rm pc}$)
as a reference position,
since the absence of the jet component 
minimize the positional change of its location.
The angular separation between Mrk~501 and NRAO~512 is $2\fdg57$.
Given the angular separation limit of dual-beam system for VERA is $2\fdg20$,
we inserted a bright calibrator 3C~345 ($z=0.593$, $1~{\rm mas}=6.64~{\rm pc}$), 
separated by $2\fdg09$ from Mrk~501.
The source configuration on the celestial sphere is shown in figure~\ref{veraalign}.
Since the flux density of 3C~345 ($\sim2$~Jy) is much higher than 
that of Mrk~501 ($\sim300$~mJy) and NRAO~512 ($\sim500$~mJy), 
we chose 3C~345 as a phase calibrator and applied its fringe phase solutions 
and structure phase solutions to the other two sources.
The two sources, 3C~345 and NRAO~512, are a well-known phase-referencing pair 
\citep{Shapiro:1979,Bartel:1986,Doi:2006,Jung:2011}.
Below, we explain our steps to derive the core position of Mrk~501 relative to NRAO~512 for each epoch.

(1) We measured 
the core position offset from the phase-tracking center on the phase-referenced images of Mrk~501
referenced to the 3C~345 core.

(2) We measured the peak position offset from the phase-tracking center on the phase-referenced images of NRAO~512
relative to 3C~345 core.

(3) We subtracted the position offset of (2) from that of (1) to derive the core position of Mrk~501 relative to NRAO~512.

Through this procedure, the position errors arise from 3C~345 were completely cancelled out (for details, see \S\ref{345result} and A.\ref{CI}).

\section{Observations}
The VERA observations were carried out 
for two sets of four adjacent days, on February 15, 16, 17, 18, and October 20, 21, 23, 24, 2011 
(details shown in table~\ref{date}).
Two sources within $2\fdg2$ were observed simultaneously by the VERA's dual-beam system \citep{Honma:2003fk}.
In the total bandwidth of 256~MHz (16 $\times$ 16~MHz),
one of the sixteen 16-MHz intermediate frequency (IF) channels was assigned to the bright calibrator, 3C~345 in A-beam.
The other 15~IF channels were allocated to the target Mrk~501, and NRAO~512 in B-beam (see also figure~\ref{veraalign}).
Left hand circular polarization (LHCP) signals were received in the Q-band,
43115-43131~MHz in A-beam and 42987-43227~MHz in B-beam.
The signals were digitized with 2-bit quantization by using the VERA digital filter \citep{Iguchi:2005qy}, 
and recorded with a rate of 1024 Mbps.

The total observation time is 8 hours per each epoch,
in which the total on source time is $\sim8$~hours for 3C~345 in A-beam,
and $\sim$6~hours for Mrk~501, $\sim$2~hours for NRAO~512 in B-beam.
The real-time instrumental phase difference data between the two beams were measured with artificial noise sources \citep{Kawaguchi:2000, Honma:2008}.
Data correlation was performed with the Mitaka FX correlator \citep{Chikada:1991}.

\section{Data reduction}~\label{datared}
Initial calibration on correlated visibilities was performed with the Astronomical Image Processing System (AIPS) software package developed by the National Radio Astronomy Observatory (NRAO).
At first, the visibilities were normalized by the auto-correlation with the AIPS task ACCOR and BPASS.
{\it A priori} amplitude calibration was performed with the AIPS task APCAL 
on the basis of the system equivalent flux density (SEFD) derived from
the opacity-corrected system temperature and the antenna gain information for each antenna.
The opacity-corrected system temperature was measured with chopper-wheel method.
We adopted the antenna gain with a correction of its dependency on the separation angle between two beams, 
based on the VERA status report in 2009\footnote{http://veraserver.mtk.nao.ac.jp/restricted/CFP2009/status09.pdf}.
The accuracy of the amplitude calibration is $\sim$10\% \citep[e.g.,][]{Nagai:2013,Petrov:2012}. 
Secondly, 
we applied the recalculated delay-tracking solutions to the correlated data \citep{Honma:2007}.
We adopted the delay-tracking center as 
($\alpha_{\rm J2000}$, $\delta_{\rm 2000}$)=(\timeform{16h53m52.216685s}, \timeform{+39D45'36.60894''}) for Mrk~501, 
($\alpha_{\rm J2000}$, $\delta_{\rm 2000}$)=(\timeform{16h42m58.809965s}, \timeform{+39D48'36.99399''}) for 3C~345, and
($\alpha_{\rm J2000}$, $\delta_{\rm 2000}$)=(\timeform{16h40m39.632772s}, \timeform{+39D46'46.02849''}) for NRAO~512,
taken from the International Celestial Reference Frame 2 (ICRF2, \citealt{Fey:2010}).
We also applied instrumental delay correction table to the data sets on B-beam 
in order to calibrate dual-beam delay difference \citep{Honma:2008}.
After that, we performed fringe fitting on the calibrator 3C~345 in the AIPS task FRING 
to solve delays, rates, and phases by assuming a point source model.
The residual phase and gain of 3C~345 were 
solved by the self-calibration imaging (described in the next paragraph).
With the AIPS task CALIB,
the phase and gain solutions of self-calibration were obtained using the source structure model.
Thirdly, we transferred the derived phase solutions of 3C~345 to NRAO~512 and Mrk~501.
After applying the amplitude solutions of Mrk~501 and NRAO~512 obtained by the self-calibration imaging,
we obtain the phase-referenced images of these two sources with the CLEAN procedure in Difmap software package \citep{Shepherd:1997}.
%

We constructed self-calibrated images of all sources.
To perform amplitude self-calibration after phase self-calibration converged, we need all the four stations.
However, since part of our observations was lacking in one or two stations due to bad weather conditions 
(see table~\ref{date}),
we combined the visibilities over four adjacent observations by AIPS task DBCON 
to increase the $uv$-coverage for obtaining precise images.
Imaging was performed in Difmap.
After careful flagging of bad visibilities and averaging over $\sim$5~seconds in the time domain, 
we constructed the initial structure models by model fitting and self-calibration iteratively.
%
%
The self-calibrated image parameters are summarized in table~\ref{selfvera}.

The peak position offsets from the phase-tracking center on the phase-referenced images of 
Mrk~501 and NRAO~512 were measured with the AIPS task JMFIT.
The peak positions measured by model-fitting to the phase-referenced visibilities
are consistent with those by JMFIT
(typically within $\sim10~\mu$as, less than $30~\mu$as)
and did not change the results significantly.
We derived their phase-referenced core positions
by subtracting from the JMFIT peak positions of the phase-referenced images
the peak position differences between JMFIT to the self-calibrated images
and model-fitting to the self-calibrated visibilities ($\sim1~\mu$as).
Then we subtracted the core position of NRAO~512 referencing to 3C~345 
from that of Mrk~501 referencing to 3C~345.
Thus, we finally obtained the core position of Mrk~501 relative to NRAO~512. 
%

\section{Results}
\subsection{Phase-referenced image qualities}
Figure~\ref{PRself} shows the phase-referenced images of Mrk~501 relative to 3C~345.
The phase-referenced images appear a point-like source due to the low SNR of the data.
We also find NRAO~512 relative to 3C~345 in the phase-referenced images
as a single-component source.
The image qualities of all the phase-referenced images are summarized in column (1)-(5) of table~\ref{vera501}.
We detect both of Mrk~501 and NRAO~512 with an image SNR of higher than nine for all epochs except for the third one, 
in which two stations suffered from bad weather condition over more than half of the observation time.
%
%
%
%
%
%
By comparing the peak fluxes of the phase-referenced images in column (3) of table~\ref{vera501} 
to those of self-calibrated images in column (2) of table~\ref{selfvera}, 
the averaged flux loss ratio is evaluated as typically $\sim20$\%, 
up to 60\% under bad weather conditions.

In appendix~\ref{veraerrsec},
the procedure to estimate the position errors is described, and the error values are summarized in table~\ref{veraerr}.
The position errors are dominated by the tropospheric zenith delay error,
by assuming typically $\sim$2~cm error for VERA \citep{Honma:2008fk}.
%
%
Since theoretically estimated thermal noise error 
(a few $\mu$as, where the image sensitivity is estimated as ~1~${\rm mJy~beam^{-1}}$ 
by assuming the system noise temperature $\sim300$~K) 
does not contribute to the random error ($\sim20~\mu$as) so much,
some other random errors and calibration errors 
still affect the quality of the phase-referenced images \citep[e.g.,][]{Doi:2006}.
Here we address them as possible as we can.
Firstly, there would be random fluctuation of tropospheric error and geometrical error.
Assuming the typical fluctuation of the propagation delay as $\sim1~{\rm cm~day^{-1}}$ \citep{Treuhaft:1987}, 
the position error would be $\sim60~\mu$as for Mrk~501-3C~345 pair,
and $\sim14~\mu$as for NRAO~512-3C~345 pair.
Secondly, the errors of self-calibration for 3C~345 
are also related to the phase-referenced image quality;
an antenna-based phase solution with a SNR of $\sim8$
corresponds to an accuracy of $7^{\circ}$ ($\sim12~\mu$as) theoretically.
Thirdly,
there would be some errors 
when we identify the core positions of Mrk~501 and NRAO~512 in the phase-referenced images
with Gaussian fitting,
because they are not point sources.
We evaluate the core identification error in the phase-referenced images as $\sim5~\mu$as,
by summing the core identification errors  ($\sim$a few~$\mu$as in A.\ref{CI})
and the Gaussian fit errors ($\sim$a few~$\mu$as) in the self-calibrated images.
However,
for most epochs, the image SNR of NRAO~512 is comparable to that of Mrk~501,
although the angular separation between NRAO~512 and 3C~345 
is one-fourth of that of Mrk~501 and 3C~345.
This means that the image SNR would not be mainly limited by
the error terms depending on the angular separations.
We would conservatively overestimate the systematic error 
(especially for the tropospheric zenith delay error).

\subsection{Mrk~501 core position relative to NRAO~512}~\label{501-512}
Figure~\ref{PRposover} shows the measured core positions of Mrk~501 relative to NRAO~512 for all the observations
as red points and blue points, 
which are listed in table~\ref{vera501512}.
%
%
The core position error of Mrk~501 relative to NRAO~512 for each epoch in table~\ref{vera501512} 
is estimated by adding the root sum square of the random error ($\sigma_{\rm random}$) 
to all the other systematic position errors of Mrk~501 and 
NRAO~512 relative to 3C~345 in table~\ref{veraerr}, except for the core identification error of 3C~345
(see \S\ref{345result}  and appendix~\ref{veraerrsec}).
%
%
Typically, the core position error of Mrk~501 relative to NRAO~512 for each epoch is
$\sim0.20~$mas in right ascension (RA) and $\sim0.21~$mas in declination (Dec),
which is around one-third of the major axis of the beam size.

The core positions of Mrk~501 relative to NRAO~512
are distributed clustering during four adjacent days in February and October 2011 in the self-calibrated image.
During the four adjacent days, 
the random error, the tropospheric error, and the ionospheric error 
($\sigma_{\rm random}$, $\sigma_{\rm trop}$, and $\sigma_{\rm ion}$ in table~\ref{veraerr})
vary randomly.
All the other errors, which are
the earth orientation parameter error, 
the antenna position error, a priori source coordinate error, 
and the core identification error 
($\sigma_{\rm earth}$, $\sigma_{\rm ant}$, $\sigma_{\rm coord}$, and $\sigma_{\rm id}$ in table~\ref{veraerr}) 
are systematic errors.
The maximum differences of the core positions within the four days
are $0.11\pm0.18~$mas and $0.08\pm0.18~$mas 
in February and October 2011, respectively.
For the errors of the maximum position differences, the common systematic errors are canceled out.
We perform a chi-squared test of the phase-referenced core positions over the four adjacent days,
and confirm that the core positions of Mrk~501 relative to NRAO~512
coincide in both RA and Dec with the significance probability of $>98\%$.
%
%
The weighted mean positions of Mrk~501's core 
relative to its phase-tracking center over the four adjacent days 
are ($x,y$) = ($-0.03\pm0.10$, $-0.05\pm0.11$)~mas in February 2011, 
and ($x,y$) = ($-0.09\pm0.10$, $0.01\pm0.10$)~mas in October 2011, 
shown as black points in figure~\ref{PRposover}.
Both of random errors and systematic errors of each epoch are used 
as the weight for calculating the weighted mean positions and their conservative errors.
The spatial distribution ranges of the core positions over the four adjacent days 
are estimated to be $\Theta\sim0.22~$mas and $\Theta\sim0.20~$mas 
in February and October 2011, respectively, with the statistical significance of $\pm1~\sigma$ 
(68\% confidence level) for the weighted mean position errors.
%
The core positions of Mrk~501 relative to NRAO~512 between February and October 2011 
(red crosses and blue crosses in figure~\ref{PRposover}) could be systematically different,
but the difference between the centers of the weighted-mean positions is $\sim0.06~$mas in both RA and Dec direction,
which is less than the spatial distribution ranges of the four adjacent days ($\sim0.20~$mas).
Therefore, we find there is no positional change of the radio core in 
Mrk~501 relative to NRAO~512 within $\sim0.20~$mas.
The discussion is shown in \S\ref{apis}.

\subsection{Core positions of Mrk~501 and NRAO~512 relative to 3C~345 core}~\label{345result}
Figures~\ref{PRpositions} (a) and (b) show the core positions of Mrk~501 and NRAO~512 relative to 3C~345 core, respectively.
The phase-referenced image qualities, the core positions and their position errors are summarized in table~\ref{vera501}. 
In the previous subsection,
we obtain the core position of Mrk~501 relative to NRAO~512 for each epoch
by subtracting the core position in figure~\ref{PRpositions} (b) 
from that in figure~\ref{PRpositions} (a).
%
Contrary to the derived position offsets 
from the phase tracking center of Mrk~501 core relative to NRAO~512 (within $\sim$0.09~mas),
the position offsets 
from the phase tracking center of Mrk~501 core relative to 3C~345 core and NRAO~512 relative to 3C~345 core
are $\sim0.4~$mas in RA and $\sim0.2~$mas in Dec,
which mainly caused by the position shifts of 3C~345 core.

\subsubsection{Mrk~501 core position relative to 3C~345 core}
The position errors of Mrk~501 core relative to 3C~345 core 
are $\sim0.12~$mas in RA and $\sim0.13~$mas in Dec, 
summarized in column (15) and (16) of table~\ref{veraerr}.
From figure~\ref{PRpositions} (a), 
the core positions of Mrk~501 relative to 3C~345 distributes clustering over the four adjacent days.
The weighted mean positions are
$(x,y)=(-0.34\pm0.06, -0.19\pm0.06)$~mas in February 2011, 
and $(x,y)=(-0.45\pm0.06, -0.20\pm0.06)$~mas in October 2011.
The weighted mean position in February 2011 is $\sim0.11~$mas east of that in October 2011,
which is less than the distribution range ($\sim0.13~$mas with $\pm1\sigma$ error).
Thus, there is no positional change of Mrk~501 core relative to 3C~345 core within $\sim0.13~$mas.
Note that the phase-referenced positions may include 
the possible positional change of Mrk~501 ($<\sim0.20~$mas) and 3C~345,
because both of them are not point sources.
Therefore, we do not discuss this result in further detail.

\subsubsection{NRAO~512 position relative to 3C~345 core}~\label{512-345}
The position errors of NRAO~512 relative to 3C~345 core
are $\sim0.04~$mas both in RA and Dec (table~\ref{veraerr}).
The estimated position errors are about one-fourth of those for Mrk~501 core relative to 3C~345 core
mainly due to the tropospheric error based on one-fourth smaller angular separation.
In figure~\ref{PRpositions} (b), 
we find the core positions of NRAO~512 relative to 3C~345 
in February 2011 (red points) and October 2011 (blue points) are clearly different.
The weighted mean positions are $(x,y)=(-0.31\pm0.02, -0.14\pm0.02)$ mas in February 2011, 
and $(x,y)=(-0.36\pm0.02, -0.21\pm0.02)$ mas in October 2011.
The weighted mean position in October 2011 is $0.05\pm0.03~$mas west and
$0.07\pm0.03~$mas south of that in February 2011.
This result indicates peak position change of 3C~345 by assuming the peak position of NRAO~512 is stationary.
The discussion is shown in \S\ref{dis512345}.

%
%
%

\section{Discussions}
\subsection{Application to internal shock model}~\label{apis}
Here we attempt to constrain bulk Lorentz factors of the jet
based the result shown above.
We assume that the location of the unresolved radio emitting regions in Mrk~501's core at 43~GHz
are identical to that of the soft X-ray emitting regions
since the radio core is optically thin against SSA.
Therefore, the internal shock model can be applicable to our target source Mrk~501.\footnote{Here we 
do not apply our result to the standing shock model.
Recently the original model of the conical standing shock proposed by \cite{Marscher:2008}
has been improved considering the multi-zone in the standing shock region 
(so called "turbulent extreme multi-zone" (TEMZ) model)
by \cite{Marscher:2014},
which also can explain the variability of the soft X-ray light curve of the blazar.
For now, our result does not rule out the model because the position of the standing shock is stationary.}
%
%
The internal shock model can well
explain the various observational properties of soft X- and gamma-ray 
light-curves in blazars and thus is regarded as
one of the leading and standard models for blazar emissions
\citep[e.g.,][]{Spada:2001fj,Tanihata:2003,Guetta:2004,Kino:2004,Mimica:2004,BYottcher:2010,Joshi:2011}.
Internal shocks will occur when a faster discrete ejecta 
catches up with slower one.
Electrons are accelerated by the shocks and the non-thermal
electromagnetic waves are emitted by these relativistic electrons
\citep[e.g.,][]{Mimica:2010,Mimica:2012}.
%
%
Then, denoting the faster ejecta's Lorentz factor $(\Gamma_{\rm f})$, 
slower ejecta's Lorentz factor $(\Gamma_{\rm s})$, 
and the initial separation of the ejecta $(I_{\rm IS})$,
the de-projected distance between the internal-shock position
and the central engine $(D_{\rm IS})$ 
is given by:
\begin{eqnarray}
D_{\rm IS}&\approx&
2\frac{\Gamma_{\rm f}^{2}\Gamma_{\rm s}^{2}}{\Gamma_{\rm f}^{2}-\Gamma_{\rm s}^{2}}I_{\rm IS}=2\frac{({\Gamma_{\rm f}/\Gamma_{\rm s}})^{2}}{({\Gamma_{\rm f}/\Gamma_{\rm s}})^{2}-1}\Gamma_{\rm s}^{2}I_{\rm IS}~\label{dmin}, \nonumber
\end{eqnarray}
where $\Gamma_{\rm f}>\Gamma_{\rm s}\gg1$.
In general, 
such collisions happen repeatedly and
multiple internal-shocked regions are generated in the jet flow.
The de-projected distributed scale of the shocked regions 
is, by definition, given by
\begin{eqnarray}
\Delta D_{\rm IS}&\equiv&D_{\rm IS,max}-D_{\rm IS,min} \label{difff},
\end{eqnarray}
where $D_{\rm IS,max}$ is the largest distance between the location of the
internal-shocked region and that of the central engine, 
and $D_{\rm IS,min}$ is the closest one.

The term $\Delta D_{\rm IS}$ for Mrk~501 is directly
constrained by our
VERA astrometric observations.
In \S\ref{501-512}, we reveals that 
the weighted mean core positions of Mrk~501
over the four adjacent days in February and October 2011
were spatially distributed within 
$\Theta\sim0.22~$mas and $\Theta\sim0.20~$mas, respectively,
by assuming the radio core of NRAO~512 is stationary.
The de-projected distributed scale $\Delta D_{\rm IS}$ can be expressed as 
follows by using the lower limit of the jet viewing angle of
$\theta_{\rm j}\ge 4^{\circ}$ \citep{Giroletti:2004a}:
\begin{eqnarray}
\Delta D_{\rm IS}
&\le&2.2\times10^{4}~R_{\rm s}\left(\frac{\Theta}{0.20~{\rm mas}}\right)\left(\frac{\theta_{\rm j}}{4^{\circ}}\right)^{-1},~\label{dis}
\end{eqnarray}
%
or $\Delta D_{\rm IS}\le1.9~$pc, where 1~mas corresponds to $7.7\times10^{3}$ 
Schwarzschild radii ($R_{\rm s}$), or 1~$R_{\rm s}=8.6\times10^{-5}$ pc.
Here we set the lower limit of the central black hole mass of Mrk~501 as
$M_{\rm BH}=0.9\times10^{9}~M_{\Sol}$ in the case of the single black hole \citep{Barth:2002}.

The right-hand-side of Eq.~(\ref{difff})
has been constrained by the previous work.
We set the separation 
between the ejecta satisfies $I_{\rm IS}\ge1R_{\rm s}$,
because $1R_{\rm s}$ is the minimum dimension of the central engine \citep[e.g.,][]{Spada:2001fj}.
\cite{Tanihata:2003} suggested that 
the Lorenz factor ratio between the faster ejecta 
and the slower ejecta is $\Gamma_{\rm f}/\Gamma_{\rm s}\le1.01$ 
to reproduce the observed X-ray light curves
in the flare state.
In quiescent state, $\Gamma_{\rm f}/\Gamma_{\rm s}$ would be less than that in the flare state
because larger $\Gamma_{\rm f}/\Gamma_{\rm s}$ produce stronger internal shocks closer to the central engine
and generate larger energy than smaller $\Gamma_{\rm f}/\Gamma_{\rm s}$ do 
when we assume constant dynamical efficiency and constant $I_{\rm IS}$.
Therefore, we can assume that
$(\Gamma_{\rm f}/\Gamma_{\rm s})_{D_{\rm IS}=D_{\rm IS,max}},~(\Gamma_{\rm f}/\Gamma_{\rm s})_{D_{\rm IS}=D_{\rm IS,min}}\le1.01$
because during our observation Mrk~501 was in relatively quiescent state 
(just before the X- and $\gamma$-ray flare; \citealt{Bartoli:2012}).
%
When we assume $\Gamma_{\rm f}/\Gamma_{\rm s}$ is constant,
$D_{\rm IS,max}$ and $D_{\rm IS,min}$ only depend on the maximum value of $\Gamma_{\rm s}$ 
($\Gamma_{\rm s,max}$) and minimum value of $\Gamma_{\rm s}$ ($\Gamma_{\rm s,min}$), respectively.
Now, $\Delta D_{\rm IS}$ can be expressed as a function of only $\Gamma_{\rm s}$.
%
Regarding the minimum value of $\Gamma_{\rm s}$, 
we adopt $\Gamma_{\rm s,min}\ge8$ 
in Mrk~501 \citep[e.g.,][]{Kino:2002} based on the minimum among the references of the one-zone SED model fitting.
Using Eqs.~(\ref{difff}) and (\ref{dis}),
we finally constrain a maximum of $\Gamma_{\rm s,max}/\Gamma_{\rm s,min}$ 
by applying the maximum or minimum values of above assumptions as follows:
\begin{eqnarray}
\!\!\!\!\!\!\!\frac{\Gamma_{\rm s,max}}{\Gamma_{\rm s,min}}\le 2.1 &&\times
\left[\frac{1}{2}\left(\frac{\Delta D_{\rm IS}}{2.2\times10^{4}~R_{\rm s}}\!\right)\right. ~\nonumber\\
&&\left.\left(\!\frac{A}{51}\right)^{-1}\left(\frac{I_{\rm IS}}{1~R_{\rm s}}\right)^{-1}\left(\frac{\Gamma_{\rm s,min}}{8}\right)^{-2}
+\!1\!\right]^{1/2},~\label{gsgs} 
\end{eqnarray}
%
where 
$A\equiv(\Gamma_{\rm f}/\Gamma_{\rm s})^{2}/[(\Gamma_{\rm f}/\Gamma_{\rm s})^{2}-1]$
and $A\ge51$.
%
%
%
Thus, we find that 
the maximum-to-minimum ratio
of $\Gamma_{\rm s}$ during our four adjacent observations
are less than $2.1$,
i.e., $8\le \Gamma_{\rm s}\le17$,
including the uncertainties of 
the relevant quantities.
The derived $\Gamma_{\rm s}$ is roughly comparable to
the one estimated by broadband spectra during its quiescent state \citep[e.g.,][]{Abdo:2011b}.
%
%
%
Further observations during its flare state are encouraged to obtain larger $\Delta D_{\rm IS}$.

Using the above $\Gamma_{\rm s}$, 
we estimate $D_{\rm IS}$, 
which corresponds to the distance between the location of the central engine
and that of the internal shock, 
where corresponds to the location of the radio core.
The estimated $D_{\rm IS}$ are
$D_{\rm IS,min}\sim0.6~{\rm pc}\left(\frac{I_{\rm IS}}{1~R_{\rm s}}\right)$, 
and $D_{\rm IS,max}\sim2.5~{\rm pc}\left(\frac{I_{\rm IS}}{1~R_{\rm s}}\right)$,
and they are comparable to the distance suggested 
in other blazars \citep{Marscher:2008, Jorstad:2010, Agudo:2011, Agudo:2011a}.
%

\subsection{Indication of peak position shift between NRAO~512 and 3C~345}~\label{dis512345}
Despite the non-detection of the core position change of the target source Mrk~501,
we find a shift of the relative peak positions 
between the two calibrators, NRAO~512 and 3C~345 (\S\ref{512-345}).
The relative peak positions of the two quasars are 
clustering over four adjacent days in February and October 2011, respectively,
and the positions of the clusters are different 
(see red points and blue points in figure~\ref{PRpositions}-b). 
From \S\ref{512-345},
the weighted mean position of NRAO~512 relative to 3C~345 core in October 2011
is $0.05\pm0.03~$mas west and $0.07\pm0.03~$mas south of that in February 2011.
This may indicate a shift of 3C~345 core of $0.05\pm0.03~$mas east and $0.07\pm0.03~$mas north
from February to October 2011,
when we assume NRAO~512 is a stationary point source.
The proper motion of 3C~345 core
is estimated to be $0.12\pm0.04~{\rm mas~yr^{-1}}$, 
equivalent to apparent velocity $4.2\pm1.4c$ 
at a position angle of $36^{+24}_{-27}$~deg (measured from north to east).
\cite{Bartel:1986} performed astrometry between this phase-referencing pair mainly at 8~GHz,
and find the proper motions of the core are not significantly different from zero
with weighted least-squares fit to the nine epoch data spanned over $\sim11$ years.
However,
focusing on their first three epoch data within one year,
the positional change of 3C~345 core was observed toward similar direction.
As for the 3C~345 jet $\sim0.60$ mas away from the core,
we find the proper motion of the absolute position $\sim0.40~{\rm mas~yr^{-1}}$
(aparent velocity $\sim14c$),
which is almost consistent with the result in \cite{Bartel:1986},
toward at a position angle of about $-98^{\circ}$ (or $72^{\circ}$).
%
Although there is a small difference in the position angle of the proper motion between the jet and the core,
a shift of the 3C~345 core could be related to the formation of the new jet component,
because the large variation of the jet ejection angle has been observed with an amplitude of $\sim20^{\circ}$ \citep[e.g.,][]{Klare:2005,Schinzel:2010}.
%
This encourages further astrometric observations.

\section{Conclusion}
In order to investigate the stationarity of the radio-core position in blazars,
we have 
conducted multi-epoch astrometric VERA observations of TeV blazar Mrk~501 at 43~GHz for the first time.
%
%
Below, we summarize the main conclusions in the present work.

\begin{enumerate}
\item 
In our observation,
we achieve an accuracy of $\sim$0.20~mas
for the radio core position of Mrk~501 relative to that of NRAO~512 
for each epoch.
The relative radio-core positions of Mrk~501
over four adjacent days in February 2011
are distributed within $\sim0.22~$mas spatial scale,
and those in October 2011 are within $\sim0.20~$mas, respectively,
with $\pm1\sigma$ error (68\% confidence level) of the weighted-mean position.
Comparing the weighted-mean position of the four adjacent observations 
in February 2011 to that obtained in October 2011,
the difference between these two positions is $\sim0.07~$mas.
Summing up, we find that 
the radio-core position of Mrk~501 during our observations
does not show significant positional change within about 0.20~mas,
by assuming the radio-core position of NRAO~512 is stationary.

\item 
By assuming the standard internal shock model,
we further constrain the bulk Lorentz factors of the ejecta based
on the observational results, i.e.,
the core position stationarity within $0.20~$mas (1.9~pc de-projected).
Then we find that
the maximum-to-minimum ratio of the slower 
ejecta's Lorenz factor can be constrained to be $\Gamma_{\rm s,max}/\Gamma_{\rm s,min}\le2.1$.
The distance between the location of the radio core and that of the central black hole
is estimated as a few pc,
which is similar to the one indicated in other blazars.


\item

An indication of core position change between February and October 2011 
was found for the phase-referencing pair of 3C~345-NRAO~512.
This detection is quite encouraging for 
the future subsequent observations of blazar
core astrometry initiated in the present work.

\end{enumerate}
Further continued astrometric observations targeting large X-ray flares of TeV blazar Mrk~501 are necessary to explore the locations and their stationarity of the radio core.
Any simultaneous gamma or radio lightcurves are also interesting, 
as well as potential changes of the optical polarization position angles \citep[e.g.,][]{Marscher:2008}.

\bigskip
The VERA is operated by Mizusawa VLBI Observatory, a branch of National Astronomical Observatory of Japan.
We thank K. M. Shibata, T. Jike, and all the staff who helped operations of the VERA observations presented in this paper.
We are grateful to T. Oyama, T. Hirota, M. Kim, N. Matsumoto, M. Sato for discussing how to reduce the VERA data.
S.K. thanks the internal referee at the MPIfR, E. Ros, for his constructive comments.
We are also grateful to S. Mineshige, K. Kohno, M. Tsuboi, K. Ebisawa, T. Mizuno for many useful comments.
We thank the anonymous referee for useful comments and suggestions. 
S.K. acknowledges this research grant provided 
by the Global COE program of University of Tokyo.
Part of this work was done with the contribution of the Italian Ministry of Foreign Affairs and University and Research for the collaboration project between Italy and Japan.
This work was partially supported by Grant-in-Aid
for Scientific Research, KAKENHI 24340042 (A.D.) and 2450240 (M.K.) from the
Japan Society for the Promotion of Science (JSPS).

\clearpage
\begin{figure*}
\begin{center}
\FigureFile(120mm,120mm){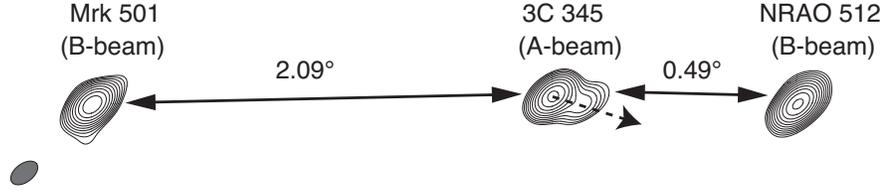}
\caption{Source configurations, dual-beam allocations, and self-calibrated images of our VERA astrometric observations at 43~GHz. The three sources quasi-perfectly align within $\sim$\timeform{3'} in east-west direction on the sky.
The calibrator 3C~345 was observed in A-beam, and at the same time, Mrk~501 and NRAO~512 were observed in B-beam in every several minutes.
Shown are the self-calibrated images of the observed sources stacked over February 15, 16, 17, 18, 2011.
The contours are plotted at $3\sigma$ noise level $\times$ (-1, 1, 2, 4,…).
The size of the restored beam (bottom left) is 0.613~mas $\times$ 0.366~mas at a position angle at $-53^{\circ}$.
The 1$\sigma$ noise level is 3.04~${\rm mJy~beam^{-1}}$ for Mrk~501, and 5.15~${\rm mJy~beam^{-1}}$ for NRAO~512.
The 1$\sigma$ noise level is 30.0~${\rm mJy~beam^{-1}}$ for 3C~345, which is about ten times higher than that for Mrk~501 because the image quality of the self-calibrated image is mainly limited by the image dynamic range (the peak flux of 3C~345 is about ten times higher than that of Mrk~501, see table~\ref{selfvera}).
Dotted arrow shows the innermost jet direction of 3C~345 at a position angle of $-103^{\circ}$.}
\label{veraalign}
\end{center}
\end{figure*}
\begin{table*}[htdp]
\caption{Summary of VERA 43~GHz astrometric observation}
\begin{center}
\begin{tabular}{ccllcc}
\hline\hline
Epoch & time range (UT) &Code & bad weather \\
\hline
2011/02/15  & 18:18-02:18 &R11046B & OGA: 0:30-1:20, 2:15-02:18, IRK: 0:50-1:40\\ 
2011/02/16 & 18:14-02:14 & R11047B & IRK, OGA: 22:00-22:40\\
2011/02/17  & 18:10-02:10 &R11048B &  IRK, ISH : 18:00-23:00\\
2011/02/18  & 18:06-02:06 & R11049B &  IRK, OGA: 23:40-23:58, 00:45-1:07\\
\hline
2011/10/20 & 01:50-09:50& R11293A &  OGA: 01:50-06:00, IRK: 06:00-09:50\\
2011/10/21 & 01:46-09:46 &R11294A & \\
2011/10/23 & 01:38-09:38&R11296A & ISH: 01:38-06:30, 05:30-07:00\\
2011/10/24 & 01:34-09:34&R11297A & \\
\hline
\end{tabular}
\end{center}
{\footnotetext ~{\bf Notes.} $-$MIZ: Mizusawa, IRK: Iriki, OGA: Ogasawara, ISH: Ishigaki. 
MIZ couldn't observe Mrk~501 for $\sim$50 min in every observation due to its elevation limit.}
\label{date}
\end{table*}

\begin{table*}[htdp]
\caption{Self-calibrated image parameters for 3C 345, Mrk~501, and NRAO~512.}
\begin{center}
\begin{tabular}{ccccccc}
\hline\hline
Source & Epoch & $I_{\rm p}$ & $\sigma_{\rm rms}$ & $I_{\rm p}/\sigma_{\rm rms}$ \\
&& ($\rm mJy~beam^{-1}$) & ($\rm mJy~beam^{-1}$) &\\
&(1) & (2) & (3) & (4)    \\
\hline
3C~345 & 2011/02 & 2210 & 30.0 & ~~74 & \\
& 2011/10 & 2040 & 17.0 & 120 & \\
\hline
Mrk~501 & 2011/02 & ~~280 & ~~~3.04 & ~~92\\
& 2011/10 & ~~239 & ~~~1.78 & 134\\
\hline
NRAO~512 & 2011/02 & ~~467 & ~~~5.15 & ~~92\\
& 2011/10 & ~~541 & ~~~1.94 & 278\\
\hline
\end{tabular}
\end{center}
{\footnotetext~ {Notes.}$-$(1) Observing epoch, (2) peak intensity, (3) rms of image noise, (4) image dynamic range.}
\label{selfvera}
\end{table*}

\begin{figure*}
\begin{center}
\FigureFile(150mm,150mm){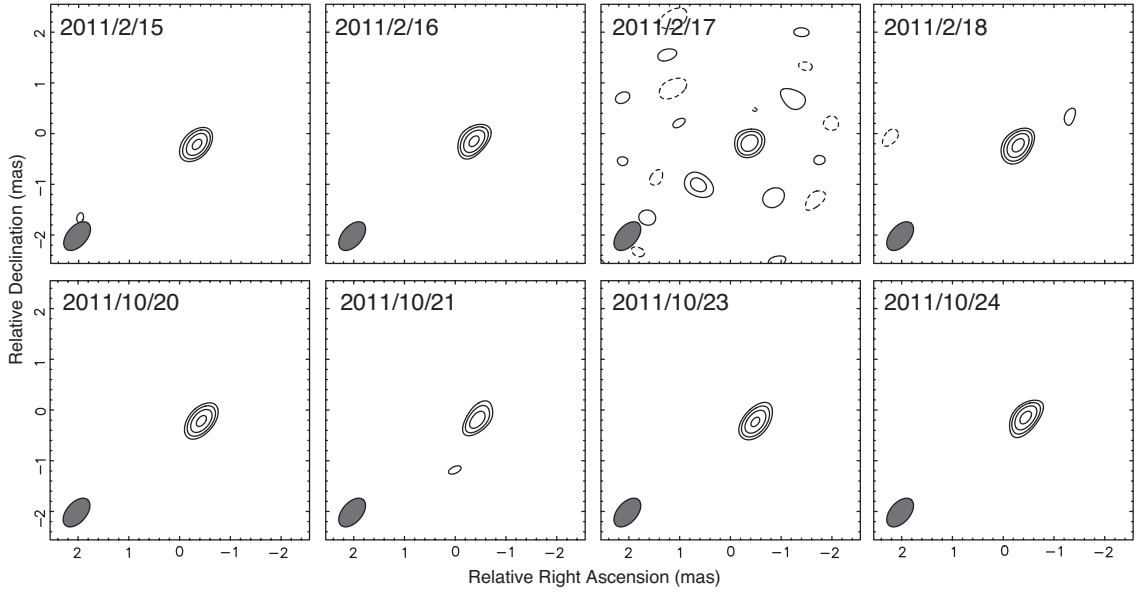}
\caption{Phase-referenced images of Mrk~501 core relative to 3C~345 core. 
The restored beam size is 0.669~mas $\times$ 0.391~mas at a position angle at $-40\fdg9$ and
the contour start from three times 28.0~${\rm mJy~beam^{-1}}$. The contours are plotted $3\sigma$ noise level $\times$ (-1, 1.41, 2, 2.82, 4). Note that the image SNR of the third epoch is much lower than that of all the others due to the bad weather condition (see table~\ref{vera501}).}
\label{PRself}
\end{center}
\end{figure*}
\begin{table*}[htdp]
\caption{Image qualities and core positions of phase-referenced images.}
\begin{center}
\begin{tabular}{ccccccc}
\hline\hline
Phase-referenced pair &Epoch  & $I_{\rm p}$ & $\sigma_{\rm rms}$ & $I_{\rm p}/\sigma_{\rm rms}$ & $\Delta\alpha$ & $\Delta\delta$ \\
&& ($\rm mJy~beam^{-1}$) & ($\rm mJy~beam^{-1}$) && (mas) & (mas)\\
(1) & (2) & (3) & (4) & (5) & (6) &(7)  \\
\hline
Mrk~501$-$3C~345&2011/02/15  & 259 & 29.1 & 9 & $-$0.33 $\pm$ 0.13 & $-$0.21 $\pm$ 0.13\\ 
&2011/02/16 & 262 & 21.0 & 12 & $-$0.37 $\pm$ 0.12 & $-$0.15 $\pm$ 0.13\\
&2011/02/17  &  224 & 44.9 & 5 & $-$0.38 $\pm$ 0.14 & -0.18 $\pm$ 0.14\\
&2011/02/18  & 275 & 30.8 & 9 & $-$0.30 $\pm$ 0.13 & -0.23 $\pm$ 0.13\\
&2011/10/20 & 262 & 15.5 & 17 & $-$0.43 $\pm$ 0.12&  -0.21 $\pm$ 0.12\\
&2011/10/21 & 210 & 24.4 & 9 & $-$0.44 $\pm$ 0.13 & -0.19 $\pm$ 0.13\\
&2011/10/23 & 254 & 18.5 & 14 & $-$0.50 $\pm$ 0.12 & -0.23$\pm$ 0.13\\
&2011/10/24 & 270 & 18.4 & 15 & $-$0.44 $\pm$ 0.12 & -0.15 $\pm$ 0.12\\
\hline
NRAO~512$-$3C~345&2011/02/15 & 388 & 28.5 & 14 & $-$0.33 $\pm$ 0.03 & $-$0.14 $\pm$ 0.04\\ 
&2011/02/16 & 408& 32.2 & 13 & $-$0.31 $\pm$ 0.03 & $-$0.15 $\pm$ 0.04\\
&2011/02/17 & 272& 61.6 & 4 & $-$0.28 $\pm$ 0.08 & $-$0.12 $\pm$ 0.07\\
&2011/02/18 & 438 & 27.5 & 16 & $-$0.30 $\pm$ 0.03 & $-$0.14 $\pm$ 0.04\\
&2011/10/20 & 228 & 17.4 & 13 & $-$0.36 $\pm$ 0.04 & $-$0.18 $\pm$ 0.04\\
&2011/10/21 & 222 & 25.2 & 9 & $-$0.32 $\pm$ 0.04 & $-$0.23 $\pm$ 0.04\\
&2011/10/23 & 424 & 38.4 & 11 & $-$0.38 $\pm$ 0.04 & $-$0.24 $\pm$ 0.04 \\
&2011/10/24 & 410 & 32.5 & 13 & $-$0.37 $\pm$ 0.04 & $-$0.18 $\pm$ 0.04 \\
\hline
\end{tabular}
\end{center}
{\footnotetext~ {Notes.}$-$(1) The first source is the target source and the second source is the reference source, (2) Observing epoch, (3) peak intensity of the target source, (4) image rms noise of the target source, (5) signal-to-noise ratio of the target source, (6) and (7) position offset from phase-tracking center of the target source.}
\label{vera501}
\end{table*}
\begin{table*}[htdp]
\caption{Estimated positional error budgets in the phase-referencing observation by VERA 43~GHz. The units are in $\mu$as.}
\begin{center} 
\small
\begin{tabular}{ccccccccccccccccc}
\hline\hline
Phase-reference pair & Epoch   & \multicolumn{2}{c}{$\sigma_{\rm random}$}  &$\sigma_{\rm trop}$ & $\sigma_{\rm ion}$  & \multicolumn{2}{c}{$\sigma_{\rm earth}$} & \multicolumn{2}{c}{$\sigma_{\rm ant}$} & \multicolumn{2}{c}{$\sigma_{\rm coord}$} & \multicolumn{2}{c}{$\sigma_{\rm id}$} &\multicolumn{2}{c}{$\sigma_{\rm rss}$}\\
(1) & (2) & (3) & (4) & (5) & (6) & (7) & (8) & (9) & (10) & (11) & (12) &(13) & (14) &(15) & (16)\\
\hline
Mrk~501$-$3C~345&2011/02/15 & 27 & 26 & 122 & $<1$ & 3 & 5 & 1 &  5 & 2 & 3 & 7 & 16 & 125 & 126 \\ 
&2011/02/16 & 16 & 16 & 122 & $<1$ & 3 & 5 & 1 &  5 & 2 & 3 & 7 & 16 & 123 & 125 \\
&2011/02/17 & 56 & 61 &122 & $<1$ & 3 & 5 & 1 &  5 & 2 & 3 & 7 & 16 &135 & 137\\
&2011/02/18 & 27 & 26 & 122 & $<1$ & 3 & 5 & 1 &  5 & 2 & 3 & 7 & 16 &125 & 126\\
&2011/10/20 & 13 & 15 &122 & $<1$ & 3 & 5 & 1 &  5 & 2 & 3 & 9 & 11 & 123 & 124\\
&2011/10/21 & 23 & 28 &122 & $<1$ & 3 & 5 & 1 &  5 & 2 & 3 & 9 & 11 & 125 & 126\\
&2011/10/23 & 16 & 21 &122 & $<1$ & 3 & 5 & 1 &  5 & 2 & 3 & 9 & 11 & 123 & 125\\
&2011/10/24 & 16 & 16 &122 & $<1$ & 3 & 5 & 1 &  5 & 2 & 3 & 9 & 11 & 123 & 124\\
\hline
NRAO~512$-$3C~345&2011/02/15 & 18 & 16 & 28 & $<1$ & 3 & 5 & 1 &  5 & 2 & 3 & 7 & 16 & 33 & 38\\ 
&2011/02/16 & 17 & 17 & 28 & $<1$ & 3 & 5 & 1 &  5 & 2 & 3 & 7 & 16 & 33 & 37\\
&2011/02/17 & 64 & 70 &28 & $<1$ & 3 & 5 & 1 &  5 & 2 & 3 & 7 & 16 & 76 & 72\\
&2011/02/18 & 15 & 14 & 28 & $<1$ & 3 & 5 & 1 &  5 & 2 & 3 & 7 & 16 & 32 & 36\\
&2011/10/20 & 17 & 18 &28 & $<1$ & 3 & 5 & 1 &  5 & 2 & 3 & 9 & 11 & 35 & 36\\
&2011/10/21 & 22 & 29 &28 & $<1$ & 3 & 5 & 1 &  5 & 2 & 3 & 9 & 11 & 42 & 38\\
&2011/10/23 & 21 & 24 &28 & $<1$ & 3 & 5 & 1 &  5 & 2 & 3 & 9 & 11 & 38 & 38\\
&2011/10/24 & 18 & 20 &28 & $<1$ & 3 & 5 & 1 &  5 & 2 & 3 & 9 & 11 & 36 & 36\\
\hline
\end{tabular}
\end{center}
{\footnotetext~ {Notes.}$-$(1) The first source is the target source and the second source is the reference source, 
(2) observing epoch, (3) and (4) the random errors estimated by the beamwidth over two times signal-to-noise ratio in RA and Dec, (5) the tropospheric residual errors, (6) the ionospheric residual errors, (7) and (8) the earth orientation parameter errors in RA and Dec, (9) and (10) the antenna position errors in RA and Dec, (11) and (12) a priori source coordinates errors in RA and Dec. The error contributions from the geometrical errors (7-12) are estimated based on the simulation presented in \cite{Pradel:2006}. (13) and (14) The core identification error of 3C~345 in RA and Dec, respectively. (15) and (16) Total errors in RA and Dec are estimated as the root-sum-square of each error.}\\
\label{veraerr}
\end{table*}

\begin{table*}[htdp]
\caption{Core positions of Mrk~501 relative to NRAO~512.}
\begin{center}
\begin{tabular}{ccccccc}
\hline\hline
Epoch &  $\Delta\alpha$ & $\Delta\delta$ \\
& (mas) & (mas)\\
(1) & (2) & (3)    \\
\hline
2011/02/15& -0.01 $\pm$ 0.20 & -0.10 $\pm$ 0.20\\
2011/02/16& -0.06 $\pm$ 0.19 & ~0.00 $\pm$ 0.20\\
2011/02/17& -0.10 $\pm$ 0.30 & -0.10 $\pm$ 0.30\\
2011/02/18& ~0.01 $\pm$ 0.19 & -0.10 $\pm$ 0.20\\
\hline
2011/10/20& -0.07 $\pm$ 0.19 & ~0.00 $\pm$ 0.20\\
2011/10/21& -0.10 $\pm$ 0.20 & ~0.00 $\pm$ 0.20\\
2011/10/23& -0.12 $\pm$ 0.19 & ~0.00 $\pm$ 0.20\\
2011/10/24& -0.07 $\pm$ 0.19 & ~0.00 $\pm$ 0.20\\
\hline
\end{tabular}
\end{center}
{\footnotetext~ {Notes.}$-$(1) Observing epoch, 
(2) and (3) core position of Mrk~501 relative to NRAO~512 in RA and Dec, respectively. 
These values are derived by subtracting the core position of Mrk~501 
relative to 3C~345 from tthat of NRAO~512 relative to 3C~345, 
summarized in (6) and (7) of table~\ref{vera501} for each epoch. 
Each position error is estimated as root-sum-square of corresponding error in table~\ref{vera501} (see \S\ref{501-512}).}\\
\label{vera501512}
\end{table*}

\begin{figure*}
  \begin{center}
   \FigureFile(60mm,60mm){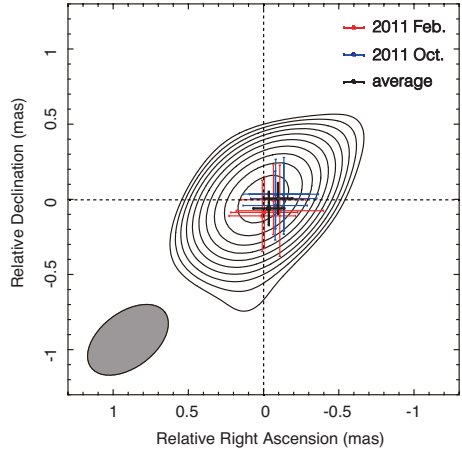}
  \end{center}
   \caption{Self-calibrated image of Mrk~501 in figure~\ref{PRself} (a) is shown with the core positions of Mrk~501 relative to NRAO~512 for all epochs overlaid. The red points represent the core positions over four adjacent days in February 2011 with an averaged positional accuracy of $0.21~$mas in RA and $0.22~$mas in Dec. The blue points show the core positions over four adjacent days in October 2011 with a positional accuracy of $0.19~$mas in RA and $0.20~$mas in Dec. Each core position is summarized in Table~\ref{vera501512}. The black points describe the weighted-mean positions over the four days in February 2011 at the relative (RA, Dec) = ($-0.03\pm0.10$, $-0.05\pm0.11$)~mas, and (RA, Dec) = ($-0.09\pm0.10$, $0.01\pm0.10$)~mas in October 2011. The origin of this figure corresponds to the phase-tracking center of Mrk~501 and the brightness peak of the self-calibrated image.}
   \label{PRposover}
\end{figure*}
\begin{figure*}
  \begin{center}
   \FigureFile(120mm,120mm){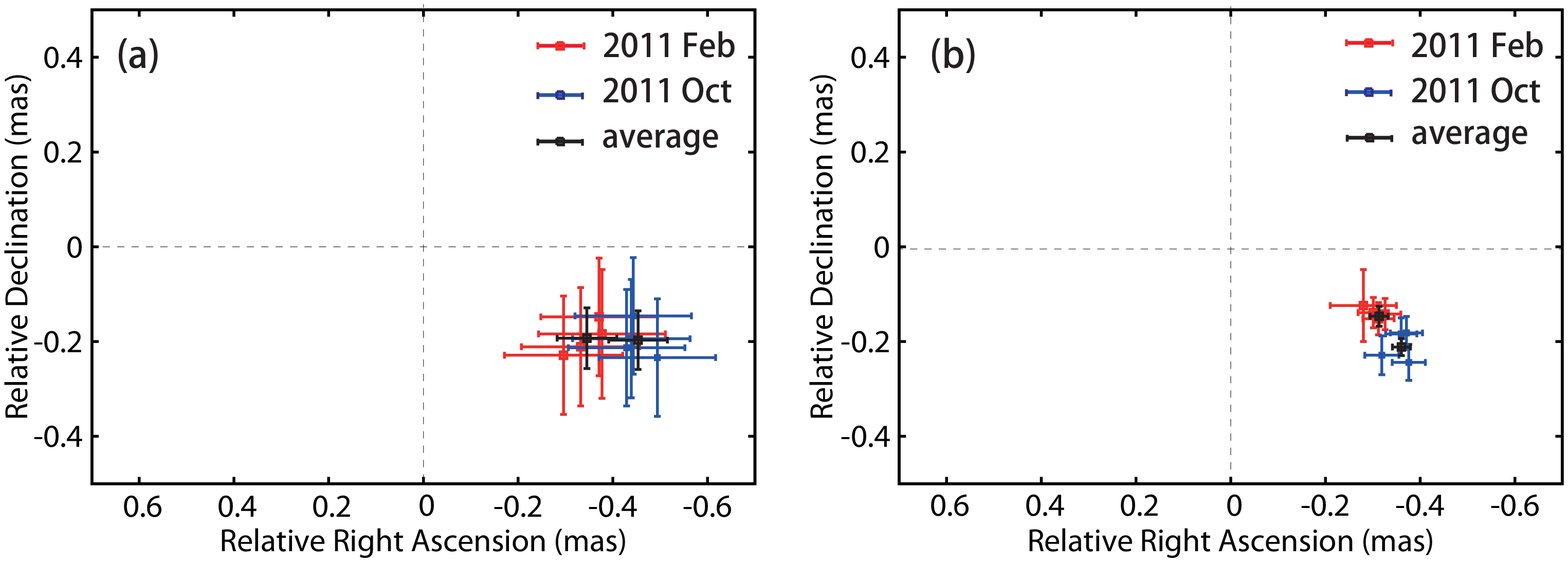}
  \end{center}
   \caption{(a) The core positions of Mrk~501 relative to 3C~345 core, and (b) those of NRAO~512 relative to 3C~345 core. 
   Red points represent four adjacent observations in February 2011, blue points represent in October 2011,
   and black points represent the weighted-mean positions over the four adjacent days (see \S\ref{345result}).}
   \label{PRpositions}
\end{figure*}
\clearpage

\begin{thebibliography}{52}
\expandafter\ifx\csname natexlab\endcsname\relax\def\natexlab#1{#1}\fi

\bibitem[{{Abdo} {et~al.}(2010){Abdo}, {Ackermann}, {Ajello}, {Axelsson},
  {Baldini}, {Ballet}, {Barbiellini}, {Bastieri}, {Baughman}, {Bechtol}, \&
  et~al.}]{Abdo:2010b}
{Abdo}, A.~A., {Ackermann}, M., {Ajello}, M., {et~al.} 2010, \nat, 463, 919

\bibitem[{{Abdo} {et~al.}(2011){Abdo}, {Ackermann}, {Ajello}, {Allafort},
  {Baldini}, {Ballet}, {Barbiellini}, {Baring}, {Bastieri}, {Bechtol}, \&
  et~al.}]{Abdo:2011b}
---. 2011, \apj, 727, 129

\bibitem[{{Agudo} {et~al.}(2011{\natexlab{a}}){Agudo}, {Jorstad}, {Marscher},
  {Larionov}, {G{\'o}mez}, {L{\"a}hteenm{\"a}ki}, {Gurwell}, {Smith},
  {Wiesemeyer}, {Thum}, {Heidt}, {Blinov}, {D'Arcangelo}, {Hagen-Thorn},
  {Morozova}, {Nieppola}, {Roca-Sogorb}, {Schmidt}, {Taylor}, {Tornikoski}, \&
  {Troitsky}}]{Agudo:2011}
{Agudo}, I., {Jorstad}, S.~G., {Marscher}, A.~P., {et~al.} 2011{\natexlab{a}},
  \apjl, 726, L13

\bibitem[{{Agudo} {et~al.}(2011{\natexlab{b}}){Agudo}, {Marscher}, {Jorstad},
  {Larionov}, {G{\'o}mez}, {L{\"a}hteenm{\"a}ki}, {Smith}, {Nilsson},
  {Readhead}, {Aller}, {Heidt}, {Gurwell}, {Thum}, {Wehrle}, {Nikolashvili},
  {Aller}, {Ben{\'{\i}}tez}, {Blinov}, {Hagen-Thorn}, {Hiriart}, {Jannuzi},
  {Joshi}, {Kimeridze}, {Kurtanidze}, {Kurtanidze}, {Lindfors}, {Molina},
  {Morozova}, {Nieppola}, {Olmstead}, {Reinthal}, {Roca-Sogorb}, {Schmidt},
  {Sigua}, {Sillanp{\"a}{\"a}}, {Takalo}, {Taylor}, {Tornikoski}, {Troitsky},
  {Zook}, \& {Wiesemeyer}}]{Agudo:2011a}
{Agudo}, I., {Marscher}, A.~P., {Jorstad}, S.~G., {et~al.} 2011{\natexlab{b}},
  \apjl, 735, L10

\bibitem[{{Bartel} {et~al.}(1986){Bartel}, {Herring}, {Ratner}, {Shapiro}, \&
  {Corey}}]{Bartel:1986}
{Bartel}, N., {Herring}, T.~A., {Ratner}, M.~I., {Shapiro}, I.~I., \& {Corey},
  B.~E. 1986, \nat, 319, 733

\bibitem[{{Barth} {et~al.}(2002){Barth}, {Ho}, \& {Sargent}}]{Barth:2002}
{Barth}, A.~J., {Ho}, L.~C., \& {Sargent}, W.~L.~W. 2002, \apjl, 566, L13

\bibitem[{{Bartoli} {et~al.}(2012){Bartoli}, {Bernardini}, {Bi}, {Bleve},
  {Bolognino}, {Branchini}, {Budano}, {Calabrese Melcarne}, {Camarri}, {Cao},
  {Cardarelli}, {Catalanotti}, {Cattaneo}, {Chen}, {Chen}, {Chen}, {Creti},
  {Cui}, {Dai}, {D'Al{\'{\i}} Staiti}, {Danzengluobu}, {Dattoli}, {De Mitri},
  {D'Ettorre Piazzoli}, {Di Girolamo}, {Ding}, {Di Sciascio}, {Feng}, {Feng},
  {Feng}, {Galeazzi}, {Giroletti}, {Gou}, {Guo}, {He}, {Hu}, {Hu}, {Huang},
  {Iacovacci}, {Iuppa}, {James}, {Jia}, {Labaciren}, {Li}, {Li}, {Li},
  {Liguori}, {Liu}, {Liu}, {Liu}, {Liu}, {Lu}, {Ma}, {Ma}, {Mancarella},
  {Mari}, {Marsella}, {Martello}, {Mastroianni}, {Montini}, {Ning}, {Pagliaro},
  {Panareo}, {Panico}, {Perrone}, {Pistilli}, {Ruggieri}, {Salvini},
  {Santonico}, {Shen}, {Sheng}, {Shi}, {Stanescu}, {Surdo}, {Tan}, {Vallania},
  {Vernetto}, {Vigorito}, {Wang}, {Wang}, {Wu}, {Wu}, {Xu}, {Xue}, {Yang},
  {Yang}, {Yao}, {Yuan}, {Zha}, {Zhang}, {Zhang}, {Zhang}, {Zhang}, {Zhang},
  {Zhang}, {Zhang}, {Zhao}, {Zhaxiciren}, {Zhaxisangzhu}, {Zhou}, {Zhu}, {Zhu},
  {Zizzi}, \& {The ARGO-YBJ Collaboration}}]{Bartoli:2012}
{Bartoli}, B., {Bernardini}, P., {Bi}, X.~J., {et~al.} 2012, \apj, 758, 2

\bibitem[{{B{\"o}ttcher} \& {Dermer}(2010)}]{BYottcher:2010}
{B{\"o}ttcher}, M., \& {Dermer}, C.~D. 2010, \apj, 711, 445

\bibitem[{{Chikada} {et~al.}(1991){Chikada}, {Kawaguchi}, {Inoue}, {Morimoto},
  {Kobayashi}, \& {Mattori}}]{Chikada:1991}
{Chikada}, Y., {Kawaguchi}, N., {Inoue}, M., {et~al.} 1991, in Frontiers of
  VLBI, ed. H.~{Hirabayashi}, M.~{Inoue}, \& H.~{Kobayashi}, 79

\bibitem[{{Doi} {et~al.}(2006){Doi}, {Fujisawa}, {Habe}, {Honma}, {Kawaguchi},
  {Kobayashi}, {Murata}, {Omodaka}, {Sudou}, \& {Takaba}}]{Doi:2006}
{Doi}, A., {Fujisawa}, K., {Habe}, A., {et~al.} 2006, \pasj, 58, 777

\bibitem[{{Fey} {et~al.}(2010)}]{Fey:2010}
{Fey}, A. L., {Gordon}, D., \& {Jacobs}, C. S. 2010, in IERS/IVS Working
Group, The Second Realization of the International Celestial Reference Frame by Very Long Baseline Interferometry, IERS Technical Note No. 35, http://www.iers.org/IERS/EN/Publications/TechnicalNotes/tn35.html

\bibitem[{{Fey} {et~al.}(2004){Fey}, {Ma}, {Arias}, {Charlot},
  {Feissel-Vernier}, {Gontier}, {Jacobs}, {Li}, \& {MacMillan}}]{Fey:2004}
{Fey}, A.~L., {Ma}, C., {Arias}, E.~F., {et~al.} 2004, \aj, 127, 3587

\bibitem[{{Giroletti} {et~al.}(2008){Giroletti}, {Giovannini}, {Cotton},
  {Taylor}, {P{\'e}rez-Torres}, {Chiaberge}, \& {Edwards}}]{Giroletti:2008}
{Giroletti}, M., {Giovannini}, G., {Cotton}, W.~D., {et~al.} 2008, \aap, 488,
  905

\bibitem[{{Giroletti} {et~al.}(2004){Giroletti}, {Giovannini}, {Taylor}, \&
  {Falomo}}]{Giroletti:2004a}
{Giroletti}, M., {Giovannini}, G., {Taylor}, G.~B., \& {Falomo}, R. 2004, \apj,
  613, 752

\bibitem[{{Guetta} {et~al.}(2004){Guetta}, {Ghisellini}, {Lazzati}, \&
  {Celotti}}]{Guetta:2004}
{Guetta}, D., {Ghisellini}, G., {Lazzati}, D., \& {Celotti}, A. 2004, \aap,
  421, 877

\bibitem[{{Hada} {et~al.}(2011){Hada}, {Doi}, {Kino}, {Nagai}, {Hagiwara}, \&
  {Kawaguchi}}]{Hada:2011}
{Hada}, K., {Doi}, A., {Kino}, M., {et~al.} 2011, \nat, 477, 185

\bibitem[{{Honma} {et~al.}(2008{\natexlab{a}}){Honma}, {Tamura}, \&
  {Reid}}]{Honma:2008fk}
{Honma}, M., {Tamura}, Y., \& {Reid}, M.~J. 2008{\natexlab{a}}, \pasj, 60, 951

\bibitem[{{Honma} {et~al.}(2003){Honma}, {Fujii}, {Hirota}, {Horiai},
  {Iwadate}, {Jike}, {Kameya}, {Kamohara}, {Kan-Ya}, {Kawaguchi}, {Kobayashi},
  {Kuji}, {Kurayama}, {Manabe}, {Miyaji}, {Nakashima}, {Omodaka}, {Oyama},
  {Sakai}, {Sakakibara}, {Sato}, {Sasao}, {Shibata}, {Shimizu}, {Suda},
  {Tamura}, {Ujihara}, \& {Yoshimura}}]{Honma:2003fk}
{Honma}, M., {Fujii}, T., {Hirota}, T., {et~al.} 2003, \pasj, 55, L57

\bibitem[{{Honma} {et~al.}(2007){Honma}, {Bushimata}, {Choi}, {Hirota}, {Imai},
  {Iwadate}, {Jike}, {Kameya}, {Kamohara}, {Kan-Ya}, {Kawaguchi}, {Kijima},
  {Kobayashi}, {Kuji}, {Kurayama}, {Manabe}, {Miyaji}, {Nagayama}, {Nakagawa},
  {Oh}, {Omodaka}, {Oyama}, {Sakai}, {Sato}, {Sasao}, {Shibata}, {Shintani},
  {Suda}, {Tamura}, {Tsushima}, \& {Yamashita Kazuyoshi}}]{Honma:2007}
{Honma}, M., {Bushimata}, T., {Choi}, Y.~K., {et~al.} 2007, \pasj, 59, 889

\bibitem[{{Honma} {et~al.}(2008{\natexlab{b}}){Honma}, {Kijima}, {Suda},
  {Kawaguchi}, {Kobayashi}, {Bushimata}, {Shimizu}, {Yoshimura}, {Sasao},
  {Hirota}, {Imai}, {Iwadate}, {Jike}, {Kameya}, {Kamohara}, {Kuji},
  {Kurayama}, {Manabe}, {Miyaji}, {Nakagawa}, {Omodaka}, {Oyama}, {Sakai},
  {Sato}, {Shibata}, \& {Tamura}}]{Honma:2008}
{Honma}, M., {Kijima}, M., {Suda}, H., {et~al.} 2008{\natexlab{b}}, \pasj, 60,
  935

\bibitem[{{Iguchi} {et~al.}(2005){Iguchi}, {Kkurayama}, {Kawaguchi}, \&
  {Kawakami}}]{Iguchi:2005qy}
{Iguchi}, S., {Kkurayama}, T., {Kawaguchi}, N., \& {Kawakami}, K. 2005, \pasj,
  57, 259

\bibitem[{{Jorstad} {et~al.}(2010){Jorstad}, {Marscher}, {Larionov}, {Agudo},
  {Smith}, {Gurwell}, {L{\"a}hteenm{\"a}ki}, {Tornikoski}, {Markowitz},
  {Arkharov}, {Blinov}, {Chatterjee}, {D'Arcangelo}, {Falcone}, {G{\'o}mez},
  {Hagen-Thorn}, {Jordan}, {Kimeridze}, {Konstantinova}, {Kopatskaya},
  {Kurtanidze}, {Larionova}, {Larionova}, {McHardy}, {Melnichuk},
  {Roca-Sogorb}, {Schmidt}, {Skiff}, {Taylor}, {Thum}, {Troitsky}, \&
  {Wiesemeyer}}]{Jorstad:2010}
{Jorstad}, S.~G., {Marscher}, A.~P., {Larionov}, V.~M., {et~al.} 2010, \apj,
  715, 362

\bibitem[{{Joshi} \& {B{\"o}ttcher}(2011)}]{Joshi:2011}
{Joshi}, M., \& {B{\"o}ttcher}, M. 2011, \apj, 727, 21

\bibitem[{{Jung} {et~al.}(2011){Jung}, {Sohn}, {Kobayashi}, {Sasao}, {Hirota},
  {Kameya}, {Choi}, \& {Chung}}]{Jung:2011}
{Jung}, T., {Sohn}, B.~W., {Kobayashi}, H., {et~al.} 2011, \pasj, 63, 375

\bibitem[{{Kawaguchi} {et~al.}(2000){Kawaguchi}, {Sasao}, \&
  {Manabe}}]{Kawaguchi:2000}
{Kawaguchi}, N., {Sasao}, T., \& {Manabe}, S. 2000, in Society of Photo-Optical
  Instrumentation Engineers (SPIE) Conference Series, Vol. 4015, Society of
  Photo-Optical Instrumentation Engineers (SPIE) Conference Series, ed. H.~R.
  {Butcher}, 544--551

\bibitem[{{Kellermann} \& {Pauliny-Toth}(1981)}]{Kellermann:1981}
{Kellermann}, K.~I., \& {Pauliny-Toth}, I.~I.~K. 1981, \araa, 19, 373

\bibitem[{{Kellermann} {et~al.}(1998){Kellermann}, {Vermeulen}, {Zensus}, \&
  {Cohen}}]{Kellermann:1998}
{Kellermann}, K.~I., {Vermeulen}, R.~C., {Zensus}, J.~A., \& {Cohen}, M.~H.
  1998, \aj, 115, 1295

\bibitem[{{Kino} {et~al.}(2004){Kino}, {Mizuta}, \& {Yamada}}]{Kino:2004}
{Kino}, M., {Mizuta}, A., \& {Yamada}, S. 2004, \apj, 611, 1021

\bibitem[{{Kino} {et~al.}(2002){Kino}, {Takahara}, \& {Kusunose}}]{Kino:2002}
{Kino}, M., {Takahara}, F., \& {Kusunose}, M. 2002, \apj, 564, 97

\bibitem[{{Klare} {et~al.}(2005){Klare}, {Zensus}, {Lobanov}, {Ros},
  {Krichbaum}, \& {Witzel}}]{Klare:2005}
{Klare}, J., {Zensus}, J.~A., {Lobanov}, A.~P., {et~al.} 2005, in Astronomical
  Society of the Pacific Conference Series, Vol. 340, Future Directions in High
  Resolution Astronomy, ed. J.~{Romney} \& M.~{Reid}, 40

\bibitem[{{Komatsu} {et~al.}(2009){Komatsu}, {Dunkley}, {Nolta}, {Bennett},
  {Gold}, {Hinshaw}, {Jarosik}, {Larson}, {Limon}, {Page}, {Spergel},
  {Halpern}, {Hill}, {Kogut}, {Meyer}, {Tucker}, {Weiland}, {Wollack}, \&
  {Wright}}]{Komatsu:2009}
{Komatsu}, E., {Dunkley}, J., {Nolta}, M.~R., {et~al.} 2009, \apjs, 180, 330

\bibitem[{{Lanyi} {et~al.}(2010){Lanyi}, {Boboltz}, {Charlot}, {Fey},
  {Fomalont}, {Geldzahler}, {Gordon}, {Jacobs}, {Ma}, {Naudet}, {Romney},
  {Sovers}, \& {Zhang}}]{Lanyi:2010}
{Lanyi}, G.~E., {Boboltz}, D.~A., {Charlot}, P., {et~al.} 2010, \aj, 139, 1695

\bibitem[{{Ma} {et~al.}(1998){Ma}, {Arias}, {Eubanks}, {Fey}, {Gontier},
  {Jacobs}, {Sovers}, {Archinal}, \& {Charlot}}]{Ma:1998}
{Ma}, C., {Arias}, E.~F., {Eubanks}, T.~M., {et~al.} 1998, \aj, 116, 516

\bibitem[{{Marscher}(2014)}]{Marscher:2014}
{Marscher}, A.~P. 2014, \apj, 780, 87

\bibitem[{{Marscher} {et~al.}(2008){Marscher}, {Jorstad}, {D'Arcangelo},
  {Smith}, {Williams}, {Larionov}, {Oh}, {Olmstead}, {Aller}, {Aller},
  {McHardy}, {L{\"a}hteenm{\"a}ki}, {Tornikoski}, {Valtaoja}, {Hagen-Thorn},
  {Kopatskaya}, {Gear}, {Tosti}, {Kurtanidze}, {Nikolashvili}, {Sigua},
  {Miller}, \& {Ryle}}]{Marscher:2008}
{Marscher}, A.~P., {Jorstad}, S.~G., {D'Arcangelo}, F.~D., {et~al.} 2008, \nat,
  452, 966

\bibitem[{{Marscher} \& {Gear}(1985)}]{Marscher:1985}
{Marscher}, A.~P., \& {Gear}, W.~K. 1985, \apj, 298, 114

\bibitem[{{Mimica} \& {Aloy}(2010)}]{Mimica:2010}
{Mimica}, P., \& {Aloy}, M.~A. 2010, \mnras, 401, 525

\bibitem[{{Mimica} \& {Aloy}(2012)}]{Mimica:2012}
---. 2012, \mnras, 421, 2635

\bibitem[{{Mimica} {et~al.}(2004){Mimica}, {Aloy}, {M{\"u}ller}, \&
  {Brinkmann}}]{Mimica:2004}
{Mimica}, P., {Aloy}, M.~A., {M{\"u}ller}, E., \& {Brinkmann}, W. 2004, \aap,
  418, 947

\bibitem[{{Nagai} {et~al.}(2013){Nagai}, {Kino}, {Niinuma}, {Akiyama}, {Hada},
  {Koyama}, {Orienti}, {Hiura}, {Sawada-Satoh}, {Honma}, {Giovannini},
  {Giroletti}, {Shibata}, \& {Sorai}}]{Nagai:2013}
{Nagai}, H., {Kino}, M., {Niinuma}, K., {et~al.} 2013, \pasj, 65, 24

\bibitem[{{Petrov} {et~al.}(2012){Petrov}, {Honma}, \& {Shibata}}]{Petrov:2012}
{Petrov}, L., {Honma}, M., \& {Shibata}, S.~M. 2012, \aj, 143, 35

\bibitem[{{Pradel} {et~al.}(2006){Pradel}, {Charlot}, \&
  {Lestrade}}]{Pradel:2006}
{Pradel}, N., {Charlot}, P., \& {Lestrade}, J.-F. 2006, \aap, 452, 1099

\bibitem[{{Reid} {et~al.}(1999){Reid}, {Readhead}, {Vermeulen}, \&
  {Treuhaft}}]{Reid:1999}
{Reid}, M.~J., {Readhead}, A.~C.~S., {Vermeulen}, R.~C., \& {Treuhaft}, R.~N.
  1999, \apj, 524, 816

\bibitem[{{Ros} {et~al.}(1999){Ros}, {Marcaide}, {Guirado}, {Ratner},
  {Shapiro}, {Krichbaum}, {Witzel}, \& {Preston}}]{Ros:1999}
{Ros}, E., {Marcaide}, J.~M., {Guirado}, J.~C., {et~al.} 1999, \aap, 348, 381

\bibitem[{{Schinzel} {et~al.}(2010){Schinzel}, {Lobanov}, \&
  {Zensus}}]{Schinzel:2010}
{Schinzel}, F.~K., {Lobanov}, A.~P., \& {Zensus}, J.~A. 2010, in Astronomical
  Society of the Pacific Conference Series, Vol. 427, Accretion and Ejection in
  AGN: a Global View, ed. L.~{Maraschi}, G.~{Ghisellini}, R.~{Della Ceca}, \&
  F.~{Tavecchio}, 153

\bibitem[{{Shapiro} {et~al.}(1979){Shapiro}, {Wittels}, {Counselman},
  {Robertson}, {Whitney}, {Hinteregger}, {Knight}, {Rogers}, {Clark}, {Hutton},
  \& {Niell}}]{Shapiro:1979}
{Shapiro}, I.~I., {Wittels}, J.~J., {Counselman}, III, C.~C., {et~al.} 1979,
  \aj, 84, 1459

\bibitem[{{Shepherd}(1997)}]{Shepherd:1997}
{Shepherd}, M.~C. 1997, in Astronomical Society of the Pacific Conference
  Series, Vol. 125, Astronomical Data Analysis Software and Systems VI, ed.
  G.~{Hunt} \& H.~{Payne}, 77

\bibitem[{{Spada} {et~al.}(2001){Spada}, {Ghisellini}, {Lazzati}, \&
  {Celotti}}]{Spada:2001fj}
{Spada}, M., {Ghisellini}, G., {Lazzati}, D., \& {Celotti}, A. 2001, \mnras,
  325, 1559

\bibitem[{{Tanihata} {et~al.}(2003){Tanihata}, {Takahashi}, {Kataoka}, \&
  {Madejski}}]{Tanihata:2003}
{Tanihata}, C., {Takahashi}, T., {Kataoka}, J., \& {Madejski}, G.~M. 2003,
  \apj, 584, 153

\bibitem[{{Thompson} {et~al.}(2001){Thompson}, {Moran}, \&
  {Swenson}}]{Thompson:2001}
{Thompson}, A.~R., {Moran}, J.~M., \& {Swenson}, Jr., G.~W. 2001,
  {Interferometry and Synthesis in Radio Astronomy, 2nd Edition}, 316, 467-506

\bibitem[{{Titov} {et~al.}(2011){Titov}, {Lambert}, \& {Gontier}}]{Titov:2011}
{Titov}, O., {Lambert}, S.~B., \& {Gontier}, A.-M. 2011, \aap, 529, A91

\bibitem[{{Treuhaft} \& {Lanyi}(1987)}]{Treuhaft:1987}
{Treuhaft}, R.~N., \& {Lanyi}, G.~E. 1987, Radio Science, 22, 251

\end{thebibliography}

\clearpage

\appendix

\section{Error Estimation}~\label{veraerrsec}
In this section, we estimate the position errors of the phase-referenced images referencing the error estimations adopted by previous VLBI astrometric studies \citep[e.g.,][]{Hada:2011,Reid:1999,Ros:1999}.
The error terms are the random errors and the uncertainties of ionospheric residuals,  tropospheric residuals, core identification process, phase-connection process, instrumental origin, and geometrical parameters (earth orientation parameters, antenna positions and a priori source coordinates) \citep{Thompson:2001}.
%
%
In the following subsections,
details of major errors are described.
Thanks to VERA's dual-beam simultaneous phase-referencing system,
the phase-connection errors can be ignored.
%
The instrumental errors are canceled out by applying the dual-beam delay difference table 
measured with noise sources with $\sim0.1$~mm accuracy \citep{Honma:2008}.
%
%
In table~\ref{veraerr}, we summarize the estimated error budget 
for each phase-referencing pair in our observations.
We consider these errors contribute independently to each other, such that 
the total position errors for each epoch can be estimated as the root-sum-square of each error.
%

\subsection{Random}~\label{random}
In column (3) and (4) of table~\ref{veraerr}, we summarize the random error of all phase-referenced images 
in RA and Dec, respectively.
The position error can be expressed as $\theta_{\mathrm{b}}/(2\times{SNR})$ \citep{Thompson:2001}, 
where $\theta_{b}$ is each interferometric beam size in RA or Dec
($\theta_{b}\sim0.6$~mas $\times~0.4$~mas at a position angle of $\sim130^{\circ}$),
and SNR is the signal-to-noise ratio of the phase-referenced images 
(see column 5 of table~\ref{vera501}).
The SNR of the phase-referenced images achieved is $\sim$11 on average.

\subsection{Propagation delay}
Here we briefly review how to estimate the contribution of the propagation delay error to the phase-referenced position error.
The effects of the propagation delay error at each station cause a decrease in SNR of the phase-referenced images and image shift.
%
%
The position error $\delta\theta$ originated from the phase error $\Delta\phi$ is expressed as 
$\delta\theta=\Delta\phi\cdot(2\pi D_{\lambda})^{-1}$ \citep[e.g.,][]{Thompson:2001},
where $D_{\lambda}\sim D\cdot\lambda^{-1}$, $D$ is the baseline length, and $\lambda$ is the observing wavelength.
$\Delta\phi$ is estimated by $2\pi c\Delta\tau\lambda^{-1}$, 
where $c$ is the speed of light and $\Delta\tau$ is the propagation delay error.
$\Delta\tau$ can be approximated by $\delta\tau_{0}~\mathrm{sec}Z$, where $\delta\tau_{0}$ is the residual vertical delay and $Z$ is the local source zenith angle.
When the two phase-referencing pair separates in zenith angle by $\Delta Z$, 
the position error $\delta\theta$ caused by the propagation delay error for a single antenna is estimated as a first-order Taylor expansion of $\tau$ as follows \citep{Reid:1999}:
\begin{eqnarray}
\delta\theta\sim\frac{c\delta\tau_{0}}{D}~\mathrm{sec}~Z~\mathrm{tan}~Z~\Delta Z. ~\label{posertau}
\end{eqnarray}

The propagation delays are mainly caused by the ionospheric and tropospheric medium.
In columns (5) and (6) of table~\ref{veraerr}, the tropospheric error and ionospheric error of each phase-referencing pair are summarized.
The propagation delays caused by the tropospheric medium are independent of observing frequency, while those caused by the ionospheric medium are inversely proportional to the square of the frequency. 
Therefore, we omitted the detail description of the ionospheric error, because it is estimated to be $<1~\mu$as at 43~GHz.
%
%
%
Since the tropospheric zenith delays are typically within $\sim$2 cm accuracy for VERA \citep{Honma:2008fk},
the position errors can be estimated as 
$\delta\theta\sim122~\mathrm{\mu as}$ for $\Delta Z=2\fdg09$ (Mrk~501-3C~345 pair), 
and $\sim28~\mathrm{\mu as}$ for $\Delta Z=0\fdg49$ (NRAO~512-3C~345 pair) 
at 43 GHz, $Z\sim50^{\circ}$ and $D=2.3\times10^{3}$~m.
%

\subsection{Core identification}~\label{CI}
The core identification error indicates the possible slight difference 
between the fitted gaussian peak 
and the actual brightness peak,
and would come from the blending of core and jet structure.
%
%
To clarify the source structure,
we produced the self-calibrated images of all the three sources (see figure~\ref{veraalign} and \S\ref{datared}).

We identified the core of 3C~345 by model fitting to the calibrated visibility data, and found it has the bright jet structure.
The core was well modeled by an elliptical Gaussian model, and the jet emission was modeled as one or two circular Gaussians.
We defined the center of each Gaussian model as the component position.
%
To evaluate the core identification error, 
we derived the differences between the center of the elliptical Gaussian and 
the position of the brightness-peak pixels in super-resolution maps measured by the AIPS task MAXFIT.
The super-resolution maps were convolved by a circular Gaussian beam of full-width at half-maximum (FWHM) 
about a half of the minor axes of the synthesized beams.
The maximum core identification error was estimated as 16~$\mu$as, 
and all the position errors are summarized in column (13) and (14) of table~\ref{veraerr}.
%
%
%
%
However, by subtracting the core position of NRAO~512 relative to 3C~345 
from the core position of Mrk~501 relative to 3C~345, 
the core identification error of calibrator 3C~345 is 
completely cancelled out.

%
For Mrk~501 and NRAO~512, 
we estimated the core identification error in the same manner.
The position differences between visibility-based model-fitting and MAXFIT
were negligible, typically less than a few $\mu$as.
%
%

\end{document}